\documentclass[twocolumn,pra,aps,showpacs]{revtex4}
\usepackage{xspace,amsmath,amsfonts,amsthm,amssymb,amsbsy,graphicx,color}

\begin{document}

\newtheorem{theo}{Theorem} 
\newtheorem{lemma}{Lemma} 

\title{The Second Law of Thermodynamics and Quantum Feedback Control: \\ Maxwell's Demon with Weak Measurements}

\author{Kurt Jacobs} 

\affiliation{Department of Physics, University of Massachusetts at Boston,
Boston, MA 02125, USA}

\begin{abstract} 
Recently Sagawa and Ueda [Phys.\ Rev.\ Lett.\ {\bf 100}, 080403 (2008)] derived a bound on the work that can be extracted from a quantum system with the use of feedback control. They left open the question of whether this bound could be achieved for every measurement that could be made by the controller. We show that it can, and that this follows straightforwardly from recent work on Maxwell's demon by Alicki {\em et al.} [Open Syst.\ Inform.\ Dynam.\ {\bf 11}, 205 (2004)], for both discrete and continuous feedback control. Our analysis also shows that bare, efficient measurements always do non-negative work on a system in equilibrium, but do not add heat. 
\end{abstract} 


\maketitle

The amount of work that can be extracted from a thermodynamical system, when it undergoes a process taking it from an initial equilibrium state $\mathcal{S}_1$ at temperature $T_1$, to a final equilibrium state $\mathcal{S}_2$ at temperature $T_2$, is given by the difference in the (Helmholtz) free energy, $F$, between these states, where 
\begin{equation}
   F \equiv E - TS , 
\end{equation}
with $E$, $T$ and $S$ being, respectively, the average internal energy, temperature and entropy of the system. For quantum systems the entropy is the von Neumann entropy, $S = -\mbox{Tr}[\rho \ln\rho]$, where $\rho$ is the density matrix of the state. The simplest example of this is the work obtained by the (quasi-equilibrium) expansion of a gas at a fixed temperature (isothermal). In this case the internal energy of the gas remains constant, and the work done by the gas is $(S_1 - S_2)/T = F_1 - F_2 = \Delta F$~\cite{SzeTherm}. 

This relation between maximum work and free energy is true for the traditional  thermodynamic processes --- that is, ones that do not have access to the microstates of the system. If we measure the system, so as to obtain information about the underlying microstate, and perform actions based on this information, then we can extract more work. This is a process of feedback control~\cite{JacobsShabani}. Of course, in this situation, a feedback controller is merely a Maxwell demon~\cite{Bennett82, Zurek84, Leff94, Lloyd97, Vedral00, Scully01, Alicki04, Kieu04, Quan06}. Sagawa and Ueda recently showed that the amount of extra work that can be extracted by a feedback controller, over and above the free-energy difference, is bounded by a measure of the information extracted by the controller~\cite{Sagawa08}. In their analysis this bound was only achieved for certain special classes of measurements (e.g. von Nuemann  measurements). Here we show, by generalizing the protocol of Alicki {\em et al.}~\cite{Alicki04}, that feedback controllers can always saturate the bound, irrespective of the measurement they make. 

The following analysis is divided into three parts. The first part introduces some important thermodynamic definitions. In the second part we derive the relationship between free energy and work extraction with feedback control, building upon previous results by Alicki {\em et al.}~\cite{Alicki04} on Maxwell's demon. Lastly, we treat the feedback controller as purely quantum mechanical, eliminating the use of quantum measurement theory, to show that the second law is preserved by the control process, in agreement with Landauer's erasure principle~\cite{Landauer61, Zurek84, Piechocinska00, Shizume95, MyEprint, Plenio01}. 

{\em Quantum Mechanics, Work, and Heat}: The average energy of a quantum system is given by $E = \mbox{Tr}[H\rho]$, where $H$ is the Hamiltonian. We can therefore write $dE = \mbox{Tr}[dH\rho] + \mbox{Tr}[Hd\rho]$. In the past the work done on the system has been equated with the first term,  $dW \equiv \mbox{Tr}[dH\rho]$, and the heat entering the system with the second: $dQ \equiv \mbox{Tr}[Hd\rho]$~\cite{Alicki04, Kieu04}. These identifications are not subtle enough for our analysis here, however, because of transformations induced by the measurement. This necessitates splitting $d\rho$ into a number of parts. Transformations that change the eigenbasis of $\rho$, such as  unitary operations, and preserve the populations of the eigenstates (and thus the entropy of the system) correspond to work done on or by the system. Alternatively, processes that leave the eigenbasis of $\rho$ fixed, but change the populations correspond to adding or subtracting heat, or information extraction.  

To examine the work done on a system by a quantum measurement, note first that the transformation of $\rho$ caused by any efficient measurement can be written as 
\begin{equation} 
   \rho \rightarrow  A \rho A^\dagger / \mbox{Tr}[A^\dagger A \rho], 
\end{equation}
for some operator $A$. The polar decomposition theorem allows us to write $A = UP$, where $U$ is unitary, and $P$ is positive, and this allows us to break up the transformation into a unitary part (work done) and an information extraction part that reduces, on average, the entropy. The operator $U$ can also include a unitary feedback operation based upon the measurement result, and, therefore, can also be {\em undone} by the use of feedback. We will refer to measurements that have no unitary part as ``bare" measurements~\cite{Jacobs05b}. 

To fully isolate the work done by the measurement we must consider the action of a  positive operator $P$ in a little more detail. If $P$ commutes with the density matrix, the only change is to the entropy. But if $P$ does not commute with $\rho$, then the action of the measurement changes both the eigenbasis of the density matrix, doing work, and generates the entropy change (extracting information). We will show in the analysis below that the average work done on a system in thermal equilibrium by a bare measurement  is always non-negative.  

The thermal equilibrium state of a system at temperature $T$ is 
\begin{equation} 
  \rho_{T} =  \frac{e^{-H/(kT)}}{\mbox{Tr}\left[ e^{-H/(kT)} \right] } , 
\end{equation} 
where $H$ is the system Hamiltonian. This state captures the fundamental assumption of statistical mechanics (that all accessible microstates are equally likely). Given this assumption, it tells us how the entropy will change with other quantities in a quasi-equilibrium process. 

{\em Extracting Work with Feedback Control:} To obtain the amount of work that can be extracted by a feedback controller, we will start our system in an equilibrium state $\rho_{T}$, denoting the initial  average energy by $E$ and the initial entropy by $S$. The first action of the controller is to measure the state, using an arbitrary quantum measurement described by a set of operators $\{P_n\}$, satisfying $\sum_n P_n^2 = 1$. The operators $P_n$ are all positive, since the unitary operator associated with each measurement outcome will be determined by the feedback chosen by the controller. After the measurement result, which occurs with probablity $p_n = \mbox{Tr}[P_n^2 \rho]$, the state is 
\begin{equation}
    \rho_n =  P_n \rho P_n / p_n . 
    \label{rhon}
\end{equation}
This is no-longer an equilibrium state, but its entropy and average energy are well-defined (its temperature is not). Call its entropy $S_n$ and its average energy $E_n$. Note that in general $E_n \not= E$, because the energy will have been changed by the measurement; this is the work done on the system by the measurement process. We will return to this later. 

Now comes the first part of the feedback control process. The controller performs work (reversible operations) on the system to transform $\rho_n$ to an equilibrium state at temperature $T$. This is achieved by i) performing a unitary operation to transform the eigenbasis of $\rho_n$ to the energy eigenbasis; ii) re-ordering the populations of the energy states so that these populations decrease monotonically with increasing energy. We will denote the populations (the eigenvalues of $\rho_n$) as $\lambda_{nj}$, and the corresponding energy levels of the system as ${\varepsilon_{nj}}$; iii) adjusting the Hamiltonian so that the separations between adjacent energy levels are such as to set $\mathcal{P}_{nj}(T) \equiv \exp(-\varepsilon_{nj}/(kT))/Z = \lambda_{nj}$, where $Z = \sum_j \exp(-\varepsilon_{nj}/(kT))$ is the partition function, and we have defined $\mathcal{P}_{nj}(T)$ as the populations of the energy eigenstates required for the system to be in thermal equilibrium at temperature $T$; iv) adjusting the Hamiltonian to produce an overall energy shift of the levels so as to return the average value of the energy to the initial value, $E$. This leaves the system in thermal equilibrium at temperature $T$, since the values of the populations, $\mathcal{P}_{nj}(T)$, required for this to be true remain unchanged by the energy shift. 

The above feedback extracts net work from the system of $\Delta E_n = E_n - E$, preserves the entropy of the state, $S_n$, and leaves the state in equilibrium at temperature $T$. 

In the second part of the feedback process, the controller performs an isothermal expansion of the system (decreasing the separation of the energy levels at fixed temperature), so as to return the entropy to the initial value, $S$. This brings the system precisely back to its initial thermal state, since the energy, temperature, and entropy of the system have all returned to their initial values. The isothermal expansion extracts $\Delta W_n = T(S - S_n)$ of work from the system. The total work extracted by the feedback controller in this cycle, given the measurement result $n$, is thus $\Delta W_n = T(S - S_n) + \Delta E_n$. Of course, the important quantity is the total {\em average} work extracted by the feedback, where the average is over the possible measurement results. This is,  
\begin{equation}
   \Delta W =  T \left( S - \sum_{n} p_n S_n \right)  + \sum_n p_n \Delta E_n. 
    \label{work1} 
\end{equation} 
We now examine the second contribution, the work extracted deterministically by the feedback, $\sum_n p_n \Delta E_n$. This is simply the average increase in the energy of the system caused by the measurement, $\Delta E_{\mbox{\scriptsize meas}} = \sum_n p_n (E_n - E)$, being extracted back by the controller. If the measurement is classical, so that all the measurement operators, $P_n$, commute with the initial state $\rho_T$ (that is, the controller measures the systems energy), then the average density matrix after the measurement is the same as the initial state: $\rho_{\mbox{\scriptsize after}} = \sum_n p_n \rho_n = \rho_T$. From this it follows immediately that $\Delta E_{\mbox{\scriptsize meas}} = 0$. Thus, as expected from our previous discussion of work and energy, when the measurement does not change the eigenbasis of the density matrix, then it does not, on average, add energy to the system. When the measurement operators do not commute with $\rho_T$, then one has $S(\rho_{\mbox{\scriptsize after}}) \geq S(\rho_T)$, a result shown by Ando~\cite{Ando89}. Because the equilibrium state, $\rho_T$, is the state with the maximum entropy given a fixed value of the average energy, it follows that $E_{\mbox{\scriptsize after}} \geq E$. We therefore have 
\begin{equation}
    \Delta E_{\mbox{\scriptsize meas}} = E_{\mbox{\scriptsize after}} - E \geq 0 . 
\end{equation}
Because the controller can always extract back as work all the energy added to the system by the measurement in a closed cycle (Eq.(\ref{work1})), to preserve the second law of thermodynamics we must interpret $\Delta E_{\mbox{\scriptsize meas}}$ as work added to the system, not heat. This is consistent with our observation that the action of a positive measurement operator induces a transformation of the density matrix eigenbasis.  

The total work extracted by the controller in a single cycle is the work extracted by the feedback process, minus the work done on the system by the measurement, and is therefore $\Delta W_{\mbox{\scriptsize fb}} = T (S - \sum_n p_n S_n)$. This is for a cycle in which the system starts in equilibrium with a given free energy, and returns back to its initial state. It now follows immediately that the work extractable by a feedback controller when starting in state $\mathcal{S}_1$ with free energy $F_1$, and ending in state $\mathcal{S}_2$ with free energy $F_2$, is 
\begin{equation}
   \Delta W_{\mbox{\scriptsize fb}} =  \Delta F + T \left( S - \sum_{n} p_n S_n \right)  , 
    \label{work2} 
\end{equation} 
where $\Delta F = F_1 - F_2$. The right hand side of this equation is the upper bound derived in~\cite{Sagawa08}.  

The quantity $\Delta S_{\mbox{\scriptsize meas}} \equiv S - \sum_{n} p_n S_n$, being the average entropy reduction provided by the measurement, is always non-negative for efficient measurements, a result due to Ozawa~\cite{Ozawa86, Nielsen, FJ}. This is a key quantity in quantum feedback control even outside thermodynamical considerations~\cite{DJJ}, and reduces to the classical mutual information when the measurement is classical. 

{\em The Second Law: } The feedback control process we have just described reduces the entropy of the bath, on average, by $\Delta S_{\mbox{\scriptsize meas}}$ during the isothermal expansion of the system (the system gains this amount of entropy from the bath). Since the final entropy of the system is the same as its initial entropy, the whole process will break the second law of thermodynamics (reduce the entropy of the universe), if the entropy of the controller does not increase by at least $\Delta S_{\mbox{\scriptsize meas}}$. The simplest way to show that the entropy of the controller does increase by the required amount is to treat the controller fully quantum mechanically. This allows us to treat the whole feedback process {\em without} using quantum measurement theory. As pointed out by Wiseman, any  feedback control process based on explicit measurements (that is, with a controller whose states are classically distinguishable, and thus do not exist in superposition states) can always be implemented with a quantum controller, without any explicit measurements~\cite{wiseman-thesis}. 

We will denote the controller as $\mathcal{C}$, and the system as $\mathcal{S}$. The measurement process is completely described by a unitary operation acting on the space of both systems. The controller has $N$ states, $|n\rangle$, $n=0,\ldots, N-1$, where $N$ is the number of measurement results. The initial state of the controller is $|0\rangle$, and that of the system is, of course, $\rho_T$. A joint unitary operation correlates the systems so that the joint state becomes 
\begin{equation} 
     \rho_{\mathcal{CS}} =  \sum_{n} p_n |n\rangle\langle n| \otimes \rho_n  + \sum_{n, m\not=n} |n\rangle\langle m| \otimes \sigma_{nm} . 
     \label{jointCS}
\end{equation} 
That this is possible is guaranteed by the fact that the $\rho_n$ are given by Eq.(\ref{rhon})~\cite{mikeandike}. The $\sigma_{nm}$ are matrices with the same dimension as the $\rho_n$, but we will not require any further details about them. A second joint unitary operation now performs feedback, applying a different unitary transformation to $\mathcal{S}$ depending on the state of the controller (each state, $|n\rangle$, of the controller is the equivalent of measurement result $n$ in our previous analysis). This unitary has the form 
\begin{equation}
     U_{\mbox{\scriptsize fb}} =  \sum_{n} |n\rangle\langle n| \otimes U_n ,  
     \label{conU}
\end{equation} 
where $U_n$ acts on the system. These unitaries perform the reordering of the eigenvalues of $\rho_n$, and the change in the system Hamiltonian (the energy levels) to bring the system into a thermal equilibrium state and adjust the average energy. 

The controller then performs the final part of the feedback in which it expands $\mathcal{S}$ isothermally to extract the work. This cannot be described purely as a unitary operation, because it leaves the bath in a state of different entropy for each value of $n$. Because these different states of the bath are necessarily macroscopically distinct (they have different entropy, and are therefore macroscopically distinguishable), this fully decoheres the controller in the basis $|n\rangle$. This can be described using a unitary of the form given in Eq.(\ref{conU}) that maximally entangles the controllers basis states, $|n\rangle$, with an auxiliary system of the same size, followed by tracing over the auxiliary system. 

Now, the result of the feedback operation on each state $\rho_n$ is to transform it to a final state $\rho_n^{\mbox{\scriptsize final}}$ with entropy $S$, temperature $T$ and average energy $E$. Since the temperature and entropy of all the $\rho_n^{\mbox{\scriptsize final}}$ are the same, they have the same set of eigenvalues (the same distribution of populations). Since the average energy is also the same for all these states, they must also have the same set of energy levels. Thus for every value of $n$ (for each state $|n\rangle$ of the controller) the system has the same final Hamiltonian and the same final state, $\rho_T$. Because of this the state of the system and controller factor, and we can write the final joint state as $\rho_{\mathcal{C}}^{\mbox{\scriptsize final}}\otimes \rho_T$. Because the system is at thermal equilibrium, the joint state of the system and bath also factors. However, the state of the controller and the bath does not factor - this is because, in general, the bath transfers a different amount of entropy to the system for each value of $n$, and is thus left in a different state for each value of $n$ (as discussed above). Since the probability is $p_n$ that the state of the controller is $|n\rangle$, and since the different states of the bath are classically distinguishable, the final state of the three systems is 
\begin{equation}
    \rho_{\mbox{\scriptsize final}} = \left( \sum_{n} p_n |n\rangle \langle n| \otimes \rho_n^{\mbox{\scriptsize bath}} \right) \otimes \rho_T . 
\end{equation}
If we denote the initial entropy of the bath as $S_{\mbox{\scriptsize B}}$, the entropy of each final bath state $\rho_n^{\mbox{\scriptsize bath}}$ is $S_{\mbox{\scriptsize B}} - (S - S_n)$. The total entropy of the final state is therefore  
\begin{equation}
     S[\rho_{\mbox{\scriptsize final}}] = S(\{p_n\}) + \sum_n p_n S_n + S_{\mbox{\scriptsize B}} , 
\end{equation}
where $ S(\{p_n\}) \equiv -\sum_{n} p_n \ln p_n$ is the entropy of the distribution of measurement results. 
Since the total initial entropy of all three systems is $S + S_{\mbox{\scriptsize B}}$, the total change in the entropy of the universe for the cycle is 
\begin{equation}
     \Delta S_{\mbox{\scriptsize tot}} = S(\{p_n\}) - \Delta S_{\mbox{\scriptsize meas}}  .
\end{equation}
The second law then follows from Nielsen's result~\cite{Nielsen}, which states that for every measurement, $S(\{p_n\})$ is an upper bound on $\Delta S_{\mbox{\scriptsize meas}}$, and thus $\Delta S_{\mbox{\scriptsize tot}} \geq 0$. 

For the controller to start the cycle again, it must return to the state $|0\rangle$. To do this it simply connects itself to a fourth system with dimension $N$ in a fixed state, $|0\rangle$, performs a (unitary) swap operation between itself and this fourth system, and then dumps the fourth system into the thermal bath. This leaves the controller in state $|0\rangle$ with zero entropy, and increases the entropy of the bath by $S(\{p_n\})$. The feedback control cycle is now complete: work $\Delta W = T\Delta S_{\mbox{\scriptsize meas}}$ has been extracted, the controller and system are back in their initial states, and the entropy of the bath has increased by $\Delta S_{\mbox{\scriptsize tot}}$.  

We note that the feedback control cycle is only thermodynamically efficient (preserves the entropy of the universe on average) when $S(\{p_n\}) = \Delta S_{\mbox{\scriptsize meas}}$. This is only true if the measurement operators $P_n$ commute with $\rho_T$~\cite{Nielsen}, so that the measurement is classical. This means that the feedback controller only preserves the entropy of the universe when it makes measurements of energy. 

We have so far only explicitly considered feedback control with efficient measurements. An inefficient measurement is one in which the controller makes an efficient measurement, but throws away some information about the measurement result~\cite{mikeandike}. All inefficient measurements can be described by the set of operators $A_{nj}$, where $\sum_{nj} A_{nj}^\dagger A_{nj} = I$, and as before $n$ labels the measurement results. The final state of the system given result $n$ is $\rho_n = \sum_{j} A_{nj}\rho_T A_{nj}^\dagger/p_n$, with $p_n = \sum_j \mbox{Tr}[A_{nj}^\dagger A_{nj} \rho_T]$. With these new definitions of $\rho_n$ and $p_n$, the above analysis of the feedback cycle goes through unchanged, except that $S_n$ is not necessarily less than $S$. In this case, the ability of the controller to extract work from the system can be {\em reduced} by the measurement, rather than increased. Because of this, inefficient measurements can add heat to a system, as well as doing work. 

Lastly, we note that we have performed all our analysis with feedback from a ``single-shot" measurement. This is usually referred to as ``discrete" feedback control, to distinguish it from feedback control that uses continuous measurement~\cite{JacobsSteck06}. However, the analysis we have presented can be easily modified to derive the same result for continuous feedback control. All we have to do is observe that each step in the feedback cycle can be performed infinitesimally. (A single infinitesimal time-step of a continuous measurement is described by a measurement in which all the operators, $A_{n}$, are infinitesimally close to the identity~\cite{JacobsSteck06}.). Our results above thus apply to all feedback control, whether discrete or continuous.


\begin{thebibliography}{25}
\expandafter\ifx\csname natexlab\endcsname\relax\def\natexlab#1{#1}\fi
\expandafter\ifx\csname bibnamefont\endcsname\relax
  \def\bibnamefont#1{#1}\fi
\expandafter\ifx\csname bibfnamefont\endcsname\relax
  \def\bibfnamefont#1{#1}\fi
\expandafter\ifx\csname citenamefont\endcsname\relax
  \def\citenamefont#1{#1}\fi
\expandafter\ifx\csname url\endcsname\relax
  \def\url#1{\texttt{#1}}\fi
\expandafter\ifx\csname urlprefix\endcsname\relax\def\urlprefix{URL }\fi
\providecommand{\bibinfo}[2]{#2}
\providecommand{\eprint}[2][]{\url{#2}}

\bibitem{SzeTherm} 
 As an introduction to thermodynamics, the author recommends the lecture notes by Sze Tan, available at \textless{}http://home.comcast.net/\textasciitilde{}szemengtan/\textgreater{}. 

\bibitem{JacobsShabani}
 An introduction to quantum feedback control is given in 
\bibinfo{author}{\bibfnamefont{K.}~\bibnamefont{Jacobs}} \bibnamefont{and}
  \bibinfo{author}{\bibfnamefont{A.}~\bibnamefont{Shabani}},
  \bibinfo{journal}{Contemp. Phys. {\bf 49}}, \bibinfo{pages}{435}  (\bibinfo{year}{2008}).

\bibitem[{\citenamefont{Bennett}(1982)}]{Bennett82}
\bibinfo{author}{\bibfnamefont{C.~H.} \bibnamefont{Bennett}},
  \bibinfo{journal}{Int. J. Theor. Phys.} \textbf{\bibinfo{volume}{21}},
  \bibinfo{pages}{905} (\bibinfo{year}{1982}).

\bibitem[{\citenamefont{Zurek}(1984)}]{Zurek84}
\bibinfo{author}{\bibfnamefont{W.~H.} \bibnamefont{Zurek}}, in
  \emph{\bibinfo{booktitle}{Frontiers of Non-Equilibrium Statistical Physics}},
  edited by \bibinfo{editor}{\bibfnamefont{G.~T.} \bibnamefont{More}}
  \bibnamefont{and} \bibinfo{editor}{\bibfnamefont{M.~O.} \bibnamefont{Scully}}
  (\bibinfo{publisher}{Plenum Press}, \bibinfo{address}{New York},
  \bibinfo{year}{1984}).

\bibitem[{\citenamefont{Leff and Rex}(1994)}]{Leff94}
\bibinfo{author}{\bibfnamefont{H.~S.} \bibnamefont{Leff}} \bibnamefont{and}
  \bibinfo{author}{\bibfnamefont{A.~F.} \bibnamefont{Rex}},
  \bibinfo{journal}{Am. J. Phys.} \textbf{\bibinfo{volume}{62}},
  \bibinfo{pages}{994} (\bibinfo{year}{1994}).

\bibitem[{\citenamefont{Ando}(1989)}]{Lloyd97}
\bibinfo{author}{\bibfnamefont{T.}~\bibnamefont{Ando}},
  \bibinfo{journal}{Linear Algebr. Appl.} \textbf{\bibinfo{volume}{118}},
  \bibinfo{pages}{163} (\bibinfo{year}{1989}).

\bibitem[{\citenamefont{Vedral}(2000)}]{Vedral00}
\bibinfo{author}{\bibfnamefont{V.}~\bibnamefont{Vedral}},
  \bibinfo{journal}{Proc. R. Soc. Lond. A} \textbf{\bibinfo{volume}{456}},
  \bibinfo{pages}{969} (\bibinfo{year}{2000}).

\bibitem[{\citenamefont{Scully}(2001)}]{Scully01}
\bibinfo{author}{\bibfnamefont{M.~O.} \bibnamefont{Scully}},
  \bibinfo{journal}{Phys. Rev. Lett.} \textbf{\bibinfo{volume}{87}},
  \bibinfo{pages}{220601} (\bibinfo{year}{2001}).

\bibitem[{\citenamefont{Alicki et~al.}(2004)\citenamefont{Alicki, Horodecki,
  Horodecki, and Horodecki}}]{Alicki04}
\bibinfo{author}{\bibfnamefont{R.}~\bibnamefont{Alicki}},
  \bibinfo{author}{\bibfnamefont{M.}~\bibnamefont{Horodecki}},
  \bibinfo{author}{\bibfnamefont{P.}~\bibnamefont{Horodecki}},
  \bibnamefont{and}
  \bibinfo{author}{\bibfnamefont{R.}~\bibnamefont{Horodecki}},
  \bibinfo{journal}{Open Sys. \& Information Dyn.}
  \textbf{\bibinfo{volume}{11}}, \bibinfo{pages}{205} (\bibinfo{year}{2004}).

\bibitem[{\citenamefont{Kieu}(2004)}]{Kieu04}
\bibinfo{author}{\bibfnamefont{T.~D.} \bibnamefont{Kieu}},
  \bibinfo{journal}{Phys. Rev. Lett.} \textbf{\bibinfo{volume}{93}},
  \bibinfo{pages}{140403} (\bibinfo{year}{2004}).

\bibitem[{\citenamefont{Quan et~al.}(2006)\citenamefont{Quan, Wang, Liu, Sun,
  and Nori}}]{Quan06}
\bibinfo{author}{\bibfnamefont{H.~T.} \bibnamefont{Quan}},
  \bibinfo{author}{\bibfnamefont{Y.~D.} \bibnamefont{Wang}},
  \bibinfo{author}{\bibfnamefont{Y.~X.} \bibnamefont{Liu}},
  \bibinfo{author}{\bibfnamefont{C.~P.} \bibnamefont{Sun}}, \bibnamefont{and}
  \bibinfo{author}{\bibfnamefont{F.}~\bibnamefont{Nori}},
  \bibinfo{journal}{Phys. Rev. Lett.} \textbf{\bibinfo{volume}{97}},
  \bibinfo{pages}{180402} (\bibinfo{year}{2006}).

\bibitem[{\citenamefont{Sagawa and Ueda}(2008)}]{Sagawa08}
\bibinfo{author}{\bibfnamefont{T.}~\bibnamefont{Sagawa}} \bibnamefont{and}
  \bibinfo{author}{\bibfnamefont{M.}~\bibnamefont{Ueda}},
  \bibinfo{journal}{Phys. Rev. Lett.} \textbf{\bibinfo{volume}{100}},
  \bibinfo{pages}{080403} (\bibinfo{year}{2008}).

\bibitem[{\citenamefont{Landauer}(1961)}]{Landauer61}
\bibinfo{author}{\bibfnamefont{R.}~\bibnamefont{Landauer}},
  \bibinfo{journal}{IBM Jl. Res. Develop.} \textbf{\bibinfo{volume}{5}},
  \bibinfo{pages}{183} (\bibinfo{year}{1961}).

\bibitem[{\citenamefont{Piechocinska}(2000)}]{Piechocinska00}
\bibinfo{author}{\bibfnamefont{B.}~\bibnamefont{Piechocinska}},
  \bibinfo{journal}{Phys. Rev. A} \textbf{\bibinfo{volume}{61}},
  \bibinfo{pages}{062314} (\bibinfo{year}{2000}).

\bibitem[{\citenamefont{Shizume}(1995)}]{Shizume95}
\bibinfo{author}{\bibfnamefont{K.}~\bibnamefont{Shizume}},
  \bibinfo{journal}{Phys. Rev. E} \textbf{\bibinfo{volume}{52}},
  \bibinfo{pages}{3495} (\bibinfo{year}{1995}).

\bibitem[{\citenamefont{Jacobs}(2005{\natexlab{a}})}]{MyEprint}
\bibinfo{author}{\bibfnamefont{K.}~\bibnamefont{Jacobs}},
  \bibinfo{journal}{Eprint: arXiv:quant-ph/0512105}
  (\bibinfo{year}{2005}{\natexlab{a}}).

\bibitem[{\citenamefont{Plenio and Vitelli}(2001)}]{Plenio01}
\bibinfo{author}{\bibfnamefont{M.~B.} \bibnamefont{Plenio}} \bibnamefont{and}
  \bibinfo{author}{\bibfnamefont{V.}~\bibnamefont{Vitelli}},
  \bibinfo{journal}{Contemp. Phys.} \textbf{\bibinfo{volume}{42}},
  \bibinfo{pages}{25} (\bibinfo{year}{2001}).

\bibitem[{\citenamefont{Jacobs}(2005{\natexlab{b}})}]{Jacobs05b}
\bibinfo{author}{\bibfnamefont{K.}~\bibnamefont{Jacobs}},
  \bibinfo{journal}{Phys. Rev. A} \textbf{\bibinfo{volume}{72}},
  \bibinfo{pages}{044101} (\bibinfo{year}{2005}{\natexlab{b}}).

\bibitem[{\citenamefont{Lloyd}(1997)}]{Ando89}
\bibinfo{author}{\bibfnamefont{S.}~\bibnamefont{Lloyd}},
  \bibinfo{journal}{Phys. Rev. A} \textbf{\bibinfo{volume}{56}},
  \bibinfo{pages}{3374} (\bibinfo{year}{1997}).

\bibitem[{\citenamefont{Ozawa}(1986)}]{Ozawa86}
\bibinfo{author}{\bibfnamefont{M.}~\bibnamefont{Ozawa}}, \bibinfo{journal}{J.
  Math. Phys.} \textbf{\bibinfo{volume}{27}}, \bibinfo{pages}{759}
  (\bibinfo{year}{1986}).

\bibitem[{\citenamefont{Nielsen}(2001)}]{Nielsen}
\bibinfo{author}{\bibfnamefont{M.~A.} \bibnamefont{Nielsen}},
  \bibinfo{journal}{Phys. Rev. A} \textbf{\bibinfo{volume}{63}},
  \bibinfo{pages}{022114} (\bibinfo{year}{2001}).

\bibitem[{\citenamefont{Fuchs and Jacobs}(2001)}]{FJ}
\bibinfo{author}{\bibfnamefont{C.~A.} \bibnamefont{Fuchs}} \bibnamefont{and}
  \bibinfo{author}{\bibfnamefont{K.}~\bibnamefont{Jacobs}},
  \bibinfo{journal}{Phys. Rev. A} \textbf{\bibinfo{volume}{63}},
  \bibinfo{pages}{062305} (\bibinfo{year}{2001}).

\bibitem[{\citenamefont{Doherty et~al.}(2001)\citenamefont{Doherty, Jacobs, and
  Jungman}}]{DJJ}
\bibinfo{author}{\bibfnamefont{A.~C.} \bibnamefont{Doherty}},
  \bibinfo{author}{\bibfnamefont{K.}~\bibnamefont{Jacobs}}, \bibnamefont{and}
  \bibinfo{author}{\bibfnamefont{G.}~\bibnamefont{Jungman}},
  \bibinfo{journal}{Phys. Rev. A} \textbf{\bibinfo{volume}{63}},
  \bibinfo{pages}{062306} (\bibinfo{year}{2001}).

\bibitem[{\citenamefont{Wiseman}(1994)}]{wiseman-thesis}
\bibinfo{author}{\bibfnamefont{H.~M.} \bibnamefont{Wiseman}},
  \bibinfo{type}{Ph.{D}. diss.}, \bibinfo{school}{The University of
  Queensland}, \bibinfo{address}{Brisbane} (\bibinfo{year}{1994}).

\bibitem[{\citenamefont{Nielsen and Chuang}(2000)}]{mikeandike}
\bibinfo{author}{\bibfnamefont{M.~A.} \bibnamefont{Nielsen}} \bibnamefont{and}
  \bibinfo{author}{\bibfnamefont{I.~L.} \bibnamefont{Chuang}},
  \emph{\bibinfo{title}{Quantum Computation and Quantum Information}}
  (\bibinfo{publisher}{Cambridge University Press}, \bibinfo{year}{2000}).

\bibitem[{\citenamefont{Jacobs and Steck}(2006)}]{JacobsSteck06}
\bibinfo{author}{\bibfnamefont{K.}~\bibnamefont{Jacobs}} \bibnamefont{and}
  \bibinfo{author}{\bibfnamefont{D.~A.} \bibnamefont{Steck}},
  \bibinfo{journal}{Contemp. Phys.} \textbf{\bibinfo{volume}{47}},
  \bibinfo{pages}{279} (\bibinfo{year}{2006}).

\end{thebibliography}

\end{document}